\documentclass{PoS}
 
\usepackage{graphicx}
\usepackage{dcolumn}
\usepackage{bm}

\title{Charmed Strange Mesons from Lattice QCD with Overlap Fermions}
 
\ShortTitle{Charmed strange mesons}
 
\author{$\chi$QCD Collaboration \\
 \speaker{S.J. Dong} and K.F. Liu\\
        Dept. of Physics and Astronomy, University of Kentucky, Lexington, KY 40506, USA\\
        E-mail: \email{super124@uky.edu}}
 
\abstract{
 The charmed-strange meson mases are calculated on a quenched lattice
QCD. The charm and strange quark propagators are calculated on the same lattice with 
the overlap fermion. $16^3\times 72$ lattices with Wilson gauge action at 
$\beta=6.3345$ are used. The charm and strange quark masses are determined by fitting the $J/\psi$ and
$\phi$ masses respectively. The charmed strange meson spectrum for the scalar, axial, pseudoscalar and
vector channels are calculated. They agree with experiments. In particular, we find the scalar meson mass
to be 2248(78) MeV which is in agreement with that of $D_{s0}^* (2317)$. 
}
 
\FullConference{The XXVI International Symposium on Lattice Field Theory\\
                 July 14-19 2008\\
                 College of William and Mary, USA}
\begin{document}
In 2003, BaBar Collaboration announced the discovery of a charmed strange 
meson $D_{s0}^*(2317)$ \cite{BaBar}. CLEO also reported the
observation of this particle in the same year~\cite{CLEO}. In the following plot,
we show the masses of the charmed strange mesons from the Particle Data Group, in which the
newly discovered $D_{s0}^*(2317)$ is a scalar meson and $D_{s1} (2460)$ and  $D_{s1}(2536)$ are
axial-vector mesons. \\

\begin{figure}[th]
\begin{center}
\caption{\bf The charmed strange meson spectrum from PDG} 
\vspace*{0.5cm}
\setlength{\unitlength}{0.240900pt}
\ifx\plotpoint\undefined\newsavebox{\plotpoint}\fi
\sbox{\plotpoint}{\rule[-0.200pt]{0.400pt}{0.400pt}}%
\begin{picture}(1349,809)(0,0)
\font\gnuplot=cmr10 at 10pt
\gnuplot
\sbox{\plotpoint}{\rule[-0.200pt]{0.400pt}{0.400pt}}%
\put(181.0,99.0){\rule[-0.200pt]{4.818pt}{0.400pt}}
\put(161,99){\makebox(0,0)[r]{1600}}
\put(1308.0,99.0){\rule[-0.200pt]{4.818pt}{0.400pt}}
\put(181.0,217.0){\rule[-0.200pt]{4.818pt}{0.400pt}}
\put(161,217){\makebox(0,0)[r]{1800}}
\put(1308.0,217.0){\rule[-0.200pt]{4.818pt}{0.400pt}}
\put(181.0,336.0){\rule[-0.200pt]{4.818pt}{0.400pt}}
\put(161,336){\makebox(0,0)[r]{2000}}
\put(1308.0,336.0){\rule[-0.200pt]{4.818pt}{0.400pt}}
\put(181.0,454.0){\rule[-0.200pt]{4.818pt}{0.400pt}}
\put(161,454){\makebox(0,0)[r]{2200}}
\put(1308.0,454.0){\rule[-0.200pt]{4.818pt}{0.400pt}}
\put(181.0,572.0){\rule[-0.200pt]{4.818pt}{0.400pt}}
\put(161,572){\makebox(0,0)[r]{2400}}
\put(1308.0,572.0){\rule[-0.200pt]{4.818pt}{0.400pt}}
\put(181.0,691.0){\rule[-0.200pt]{4.818pt}{0.400pt}}
\put(161,691){\makebox(0,0)[r]{2600}}
\put(1308.0,691.0){\rule[-0.200pt]{4.818pt}{0.400pt}}
\put(181.0,809.0){\rule[-0.200pt]{4.818pt}{0.400pt}}
\put(161,809){\makebox(0,0)[r]{2800}}
\put(40,424){\makebox(0,0){\rotatebox{90}{Mass (MeV)}}}
\put(304,-18){\makebox(0,0){\small $D_{s2}(2^+)$}}
\put(460,-18){\makebox(0,0){\small $D_s(0^{-})$}}
\put(640,-18){\makebox(0,0){\small $D_{s0}^*(0^+)$}}
\put(795,-18){\makebox(0,0){\small $D_{s1}(1^+)$}}
\put(959,-18){\makebox(0,0){\small $D_{s1}(1^+)$}}
\put(1123,-18){\makebox(0,0){\small $D_s^*(1^-?)$}}
\put(304,732){\makebox(0,0){\footnotesize $D_{s2}(2573)$}}
\put(427,377){\makebox(0,0){\footnotesize $D_{s}(1969)$}}
\put(755,667){\makebox(0,0){\footnotesize $D_{s1}(2460)$}}
\put(918,714){\makebox(0,0){\footnotesize $D_{s1}(2536)$}}
\put(591,584){\makebox(0,0){\footnotesize $D_{s0}^*(2317)$}}
\put(1082,461){\makebox(0,0){\footnotesize $D_{s}^*(2112)$}}
\put(1308.0,809.0){\rule[-0.200pt]{4.818pt}{0.400pt}}
\put(263,675){\usebox{\plotpoint}}
\put(263.0,675.0){\rule[-0.200pt]{19.754pt}{0.400pt}}
\put(427,317){\usebox{\plotpoint}}
\put(427.0,317.0){\rule[-0.200pt]{19.754pt}{0.400pt}}
\put(591,524){\usebox{\plotpoint}}
\put(591.0,524.0){\rule[-0.200pt]{19.754pt}{0.400pt}}
\put(755,608){\usebox{\plotpoint}}
\put(755.0,608.0){\rule[-0.200pt]{19.513pt}{0.400pt}}
\put(918,652){\usebox{\plotpoint}}
\put(918.0,652.0){\rule[-0.200pt]{19.754pt}{0.400pt}}
\put(1082,399){\usebox{\plotpoint}}
\put(1082.0,399.0){\rule[-0.200pt]{19.754pt}{0.400pt}}
\end{picture}
\end{center}
\end{figure}
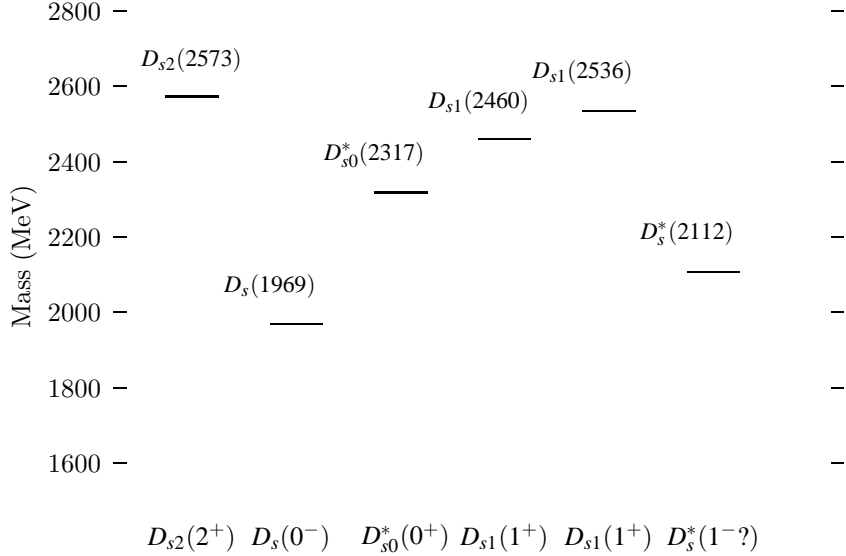

 Predictions of these charmed strange meson spectrum have been made in the quark model~\cite{Godfrey}. 
While they gave good prediction for the tensor and ${}^3P_1$ axial-vector, pseudoscalar, and
vector mesons, its prediction of the ${}^1P_1$ axial-vector at 2.53 GeV is $\sim 70$ MeV above
the experimental $D_{s1} (2460)$. More puzzling is the prediction of the scalar meson at
2.48 GeV which is $\sim 160$ MeV above $D_{s0}^* (2317)$. This discrepancy has prompted
the speculations that $D_{s0}^*(2317)$ is a DK molecule~\cite{bcl03} or four quark state~\cite{Cheng} 
or a threshold effect~\cite{br03} instead of a $c\bar{s}$ meson.

 There are also a few lattice calculations. Lattice 
NRQCD calculation with quenched approximation gives $m(D_{s0}^*)=2.44(5)$GeV \cite{Lewis}. 
The $n_f=2$ calculation with the heavy quark at the static limit gives 
$m(D_{s0}^*)=2.57(11)$GeV~\cite{Gunnar}. These are also 
significantly heavier than the experimental mass of 2.317GeV. The recent
calculation with a relativistic heavy quark ( RHQ ) action gives $\Delta m=m(D_{s0}^*)-m(D_s)
=0.1243(28)$GeV, or $m(D_{s0}^*)=2.093(3)$GeV~\cite{CP-PACS}, which is
significantly lower than the experimental mass of $D_{s0}^*$.

 Although, in principle, lattice QCD is an ideal tool to calculate hadron
spectrum from the first principle, in practice it suffers from systematic errors such as due
to discretization effects. These errors can be large for fermion actions
which do not have chiral symmetry at finite lattice spacing. In particular,
the $ma$ errors can be substantial for heavy quarks with the commonly used
lattice spacing $a \sim 0.1$ fm.

 The overlap fermion action obeys chiral symmetry at finite lattice
spacing and is, thus, free of $O(a)$ and $O(ma)$ errors. It is shown
that the effective quark propagator of the massive overlap fermion has the same
form as that of the continuum~\cite{Liu1}, i.e.
\begin{eqnarray}   \label{eff_prop}
D_{ov}(m) &=& D_{ov} + ma(1-\frac{1}{2} D_{ov}) \\
S_{eff} &=& \frac{1 -\frac{1}{2} D_{ov}}{D_{ov}(m)} =  \frac{1}{D_c + m },
\end{eqnarray}
where $D_c = \frac{D_{ov}}{1- \frac{1}{2} D_{ov}}$ and $\{D_c, \gamma_5\} = 0$.

As it can be seen from Eq.~(\ref{eff_prop}), the formalism is the same for all
quark masses. As such, it can be used for both light and heavy quarks as long as 
the $O(m^2a^2)$ errors are negligible~\cite{Liu1}. In fact, the $O(a^2)$ error is also small 
\cite{Draper}. By examining the dispersion relations and the hyperfine
splittings, it is found that one could use $ma \leq 0.5$ and still keeps the 
$O(m^2a^2)$ errors to less than 3\% to 4\% \cite{Liu2}.

We carried out a study of charmonium on a quenched $16^3\times 72$ lattice with Wilson 
gauge action at $\beta = 6.3345$. With the $r_0=0.5$fm scale we
obtain $a=0.0560$fm. Multi-mass overlap inverter is used to calculate propagators for 26 
quark masses for $ma$ from 0.02 to 0.85 \cite{Sonali}.

\begin{figure}[th]
\begin{center}
\caption{\bf Quark masses used to calculate $J/\psi$ and $\phi$ masses on the $16x72$ lattice.}
\vspace*{0.2cm}
\setlength{\unitlength}{0.240900pt}
\ifx\plotpoint\undefined\newsavebox{\plotpoint}\fi
\sbox{\plotpoint}{\rule[-0.200pt]{0.400pt}{0.400pt}}%
\begin{picture}(1200,900)(0,0)
\font\gnuplot=cmr10 at 10pt
\gnuplot
\sbox{\plotpoint}{\rule[-0.200pt]{0.400pt}{0.400pt}}%
\put(161.0,123.0){\rule[-0.200pt]{4.818pt}{0.400pt}}
\put(141,123){\makebox(0,0)[r]{1}}
\put(1119.0,123.0){\rule[-0.200pt]{4.818pt}{0.400pt}}
\put(161.0,232.0){\rule[-0.200pt]{4.818pt}{0.400pt}}
\put(141,232){\makebox(0,0)[r]{1.5}}
\put(1119.0,232.0){\rule[-0.200pt]{4.818pt}{0.400pt}}
\put(161.0,341.0){\rule[-0.200pt]{4.818pt}{0.400pt}}
\put(141,341){\makebox(0,0)[r]{2}}
\put(1119.0,341.0){\rule[-0.200pt]{4.818pt}{0.400pt}}
\put(161.0,450.0){\rule[-0.200pt]{4.818pt}{0.400pt}}
\put(141,450){\makebox(0,0)[r]{2.5}}
\put(1119.0,450.0){\rule[-0.200pt]{4.818pt}{0.400pt}}
\put(161.0,559.0){\rule[-0.200pt]{4.818pt}{0.400pt}}
\put(141,559){\makebox(0,0)[r]{3}}
\put(1119.0,559.0){\rule[-0.200pt]{4.818pt}{0.400pt}}
\put(161.0,668.0){\rule[-0.200pt]{4.818pt}{0.400pt}}
\put(141,668){\makebox(0,0)[r]{3.5}}
\put(1119.0,668.0){\rule[-0.200pt]{4.818pt}{0.400pt}}
\put(161.0,777.0){\rule[-0.200pt]{4.818pt}{0.400pt}}
\put(141,777){\makebox(0,0)[r]{4}}
\put(1119.0,777.0){\rule[-0.200pt]{4.818pt}{0.400pt}}
\put(161.0,123.0){\rule[-0.200pt]{0.400pt}{4.818pt}}
\put(161,82){\makebox(0,0){0}}
\put(161.0,757.0){\rule[-0.200pt]{0.400pt}{4.818pt}}
\put(346.0,123.0){\rule[-0.200pt]{0.400pt}{4.818pt}}
\put(346,82){\makebox(0,0){0.1}}
\put(346.0,757.0){\rule[-0.200pt]{0.400pt}{4.818pt}}
\put(530.0,123.0){\rule[-0.200pt]{0.400pt}{4.818pt}}
\put(530,82){\makebox(0,0){0.2}}
\put(530.0,757.0){\rule[-0.200pt]{0.400pt}{4.818pt}}
\put(715.0,123.0){\rule[-0.200pt]{0.400pt}{4.818pt}}
\put(715,82){\makebox(0,0){0.3}}
\put(715.0,757.0){\rule[-0.200pt]{0.400pt}{4.818pt}}
\put(899.0,123.0){\rule[-0.200pt]{0.400pt}{4.818pt}}
\put(899,82){\makebox(0,0){0.4}}
\put(899.0,757.0){\rule[-0.200pt]{0.400pt}{4.818pt}}
\put(1084.0,123.0){\rule[-0.200pt]{0.400pt}{4.818pt}}
\put(1084,82){\makebox(0,0){0.5}}
\put(1084.0,757.0){\rule[-0.200pt]{0.400pt}{4.818pt}}
\put(161.0,123.0){\rule[-0.200pt]{235.600pt}{0.400pt}}
\put(1139.0,123.0){\rule[-0.200pt]{0.400pt}{157.549pt}}
\put(161.0,777.0){\rule[-0.200pt]{235.600pt}{0.400pt}}
\put(40,450){\makebox(0,0){\rotatebox{90}{\large\bf Mass of $\phi$ and $J/\psi$ (GeV)}}}
\put(650,21){\makebox(0,0){\Large\bf $m_q a$}}
\put(650,839){\makebox(0,0){\large\bf $16^3\times 72$ with Overlap Fermions}}
\put(161.0,123.0){\rule[-0.200pt]{0.400pt}{157.549pt}}
\put(979,737){\makebox(0,0){\large\bf $\phi$}}
\put(999.0,737.0){\rule[-0.200pt]{24.090pt}{0.400pt}}
\put(999.0,727.0){\rule[-0.200pt]{0.400pt}{4.818pt}}
\put(1099.0,727.0){\rule[-0.200pt]{0.400pt}{4.818pt}}
\put(290.0,179.0){\rule[-0.200pt]{0.400pt}{0.964pt}}
\put(280.0,179.0){\rule[-0.200pt]{4.818pt}{0.400pt}}
\put(280.0,183.0){\rule[-0.200pt]{4.818pt}{0.400pt}}
\put(309.0,189.0){\rule[-0.200pt]{0.400pt}{1.204pt}}
\put(299.0,189.0){\rule[-0.200pt]{4.818pt}{0.400pt}}
\put(299.0,194.0){\rule[-0.200pt]{4.818pt}{0.400pt}}
\put(327.0,200.0){\rule[-0.200pt]{0.400pt}{0.964pt}}
\put(317.0,200.0){\rule[-0.200pt]{4.818pt}{0.400pt}}
\put(317.0,204.0){\rule[-0.200pt]{4.818pt}{0.400pt}}
\put(438.0,270.0){\rule[-0.200pt]{0.400pt}{0.964pt}}
\put(428.0,270.0){\rule[-0.200pt]{4.818pt}{0.400pt}}
\put(428.0,274.0){\rule[-0.200pt]{4.818pt}{0.400pt}}
\put(622.0,380.0){\rule[-0.200pt]{0.400pt}{0.482pt}}
\put(612.0,380.0){\rule[-0.200pt]{4.818pt}{0.400pt}}
\put(290,181){\raisebox{-.8pt}{\makebox(0,0){$\Diamond$}}}
\put(309,191){\raisebox{-.8pt}{\makebox(0,0){$\Diamond$}}}
\put(327,202){\raisebox{-.8pt}{\makebox(0,0){$\Diamond$}}}
\put(438,272){\raisebox{-.8pt}{\makebox(0,0){$\Diamond$}}}
\put(622,381){\raisebox{-.8pt}{\makebox(0,0){$\Diamond$}}}
\put(1049,737){\raisebox{-.8pt}{\makebox(0,0){$\Diamond$}}}
\put(612.0,382.0){\rule[-0.200pt]{4.818pt}{0.400pt}}
\put(979,696){\makebox(0,0){\large\bf $J/\psi$}}
\put(999.0,696.0){\rule[-0.200pt]{24.090pt}{0.400pt}}
\put(999.0,686.0){\rule[-0.200pt]{0.400pt}{4.818pt}}
\put(1099.0,686.0){\rule[-0.200pt]{0.400pt}{4.818pt}}
\put(807.0,471.0){\usebox{\plotpoint}}
\put(797.0,471.0){\rule[-0.200pt]{4.818pt}{0.400pt}}
\put(797.0,472.0){\rule[-0.200pt]{4.818pt}{0.400pt}}
\put(991.0,585.0){\usebox{\plotpoint}}
\put(981.0,585.0){\rule[-0.200pt]{4.818pt}{0.400pt}}
\put(981.0,586.0){\rule[-0.200pt]{4.818pt}{0.400pt}}
\put(1028.0,607.0){\usebox{\plotpoint}}
\put(1018.0,607.0){\rule[-0.200pt]{4.818pt}{0.400pt}}
\put(1018.0,608.0){\rule[-0.200pt]{4.818pt}{0.400pt}}
\put(1084.0,640.0){\usebox{\plotpoint}}
\put(1074.0,640.0){\rule[-0.200pt]{4.818pt}{0.400pt}}
\put(807,472){\circle*{24}}
\put(991,585){\circle*{24}}
\put(1028,608){\circle*{24}}
\put(1084,641){\circle*{24}}
\put(1049,696){\circle*{24}}
\put(1074.0,641.0){\rule[-0.200pt]{4.818pt}{0.400pt}}
\end{picture}
\end{center}
\end{figure}
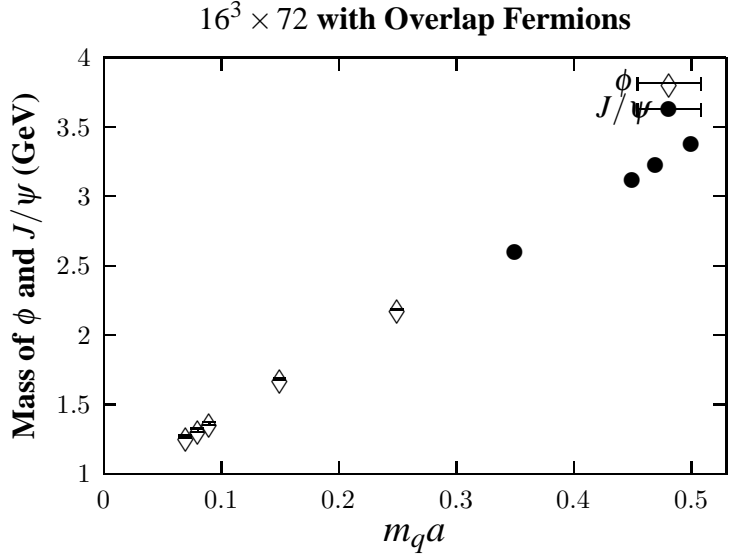
\newpage
\clearpage
 Matching the $J/\psi$ mass $m(J/\psi)=3.097$GeV, we obtain the
charm quark mass in lattice unit to be $m_c a=0.431(1)$. Inputting the
$\phi(1^-)$ particle mass $m(\phi)=1.020$GeV, we obtain the strange quark mass
in lattice unit to be $m_s a=0.0205(32)$. The vector meson mass for a range of 
quark masses is shown in Fig. 2. We find that the charm and strange 
quark masses are in the range $0.01 \leq m_q a \leq 0.5$. Thus
the $O(m^2a^2)$ error is expected to be around 1\% for the masses of the charmed strange mesons..

We construct the charmed strange meson 
correlators with the standard local interpreting fields:
\begin{eqnarray}
0^- &\Longrightarrow & \chi(x) = \bar{\psi}(x)\gamma_5 \psi(x)\\
0^+ &\Longrightarrow & \chi(x) = \bar{\psi}(x) \psi(x) \\
1^- &\Longrightarrow & \chi(x) = \bar{\psi}(x)\gamma_j \psi(x) ~~~~j=1,2,3\\
1^+ &\Longrightarrow & \chi_a(x) = \bar{\psi}(x)\gamma_5\gamma_j
\psi(x) ~~~~j=1,2,3\\
1^+ &\Longrightarrow & \chi_b(x) = \bar{\psi}(x)\gamma_i\gamma_j
\psi(x) ~~~~\{ij\}=\{12\},\{23\},\{31\}
\end{eqnarray}

We do not consider the mixing between $\chi_a$ and $\chi_b$ for the axial-vector
mesons in this work.

 All the meson correlators with 100 configurations show reasonably nice cosh behavior, 
such as in the scalar channel in Fig. 3.
 
\begin{figure}[th]
\begin{center}
\caption{\bf The charmed strange meson correlator in the scalar channel.}
\vspace*{0.1cm}
\input{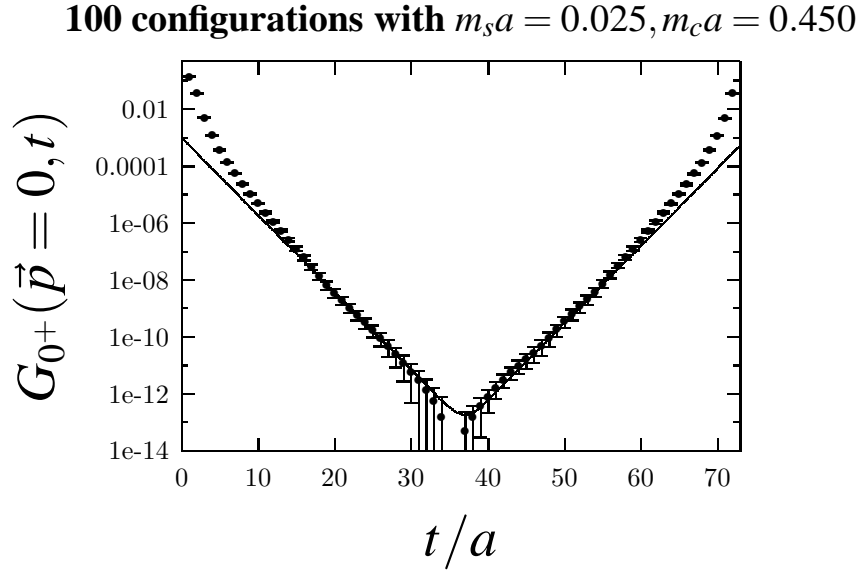}
\end{center}
\end{figure}

We used 5 quark masses each around those of the charm and strange to construct 25 correlators 
for the above interpolation fields.
We first fix the light quark mass and vary the heavy quark masses to fit the
meson masses which correspond to the charm mass at $m_c a=0.431$. The results for the
scalar and vector mesons are plotted against the light quark mass in Fig.4.

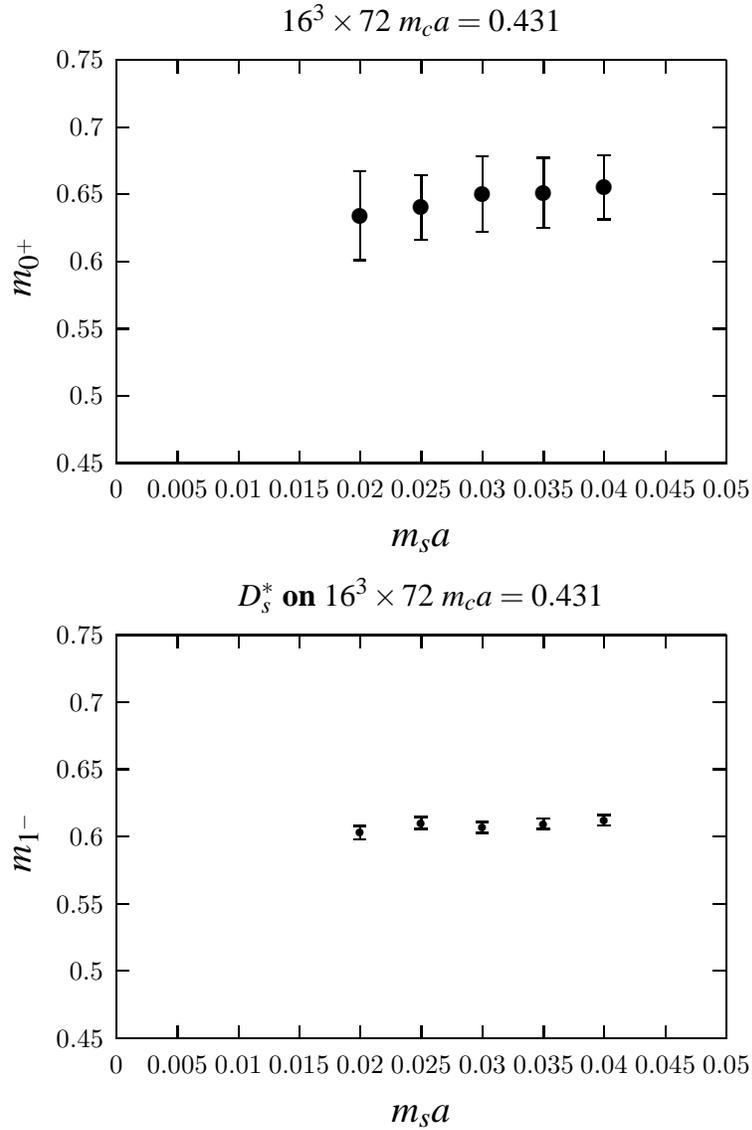
\begin{figure}[th]
\begin{center}
\caption{\bf The charmed scalar and vector 
meson mass varying with light quark mass}
\vspace*{0.1cm}
\setlength{\unitlength}{0.240900pt}
\ifx\plotpoint\undefined\newsavebox{\plotpoint}\fi
\sbox{\plotpoint}{\rule[-0.200pt]{0.400pt}{0.400pt}}%
\begin{picture}(1200,900)(0,0)
\font\gnuplot=cmr10 at 10pt
\gnuplot
\sbox{\plotpoint}{\rule[-0.200pt]{0.400pt}{0.400pt}}%
\put(181.0,143.0){\rule[-0.200pt]{4.818pt}{0.400pt}}
\put(161,143){\makebox(0,0)[r]{0.45}}
\put(1119.0,143.0){\rule[-0.200pt]{4.818pt}{0.400pt}}
\put(181.0,249.0){\rule[-0.200pt]{4.818pt}{0.400pt}}
\put(161,249){\makebox(0,0)[r]{0.5}}
\put(1119.0,249.0){\rule[-0.200pt]{4.818pt}{0.400pt}}
\put(181.0,354.0){\rule[-0.200pt]{4.818pt}{0.400pt}}
\put(161,354){\makebox(0,0)[r]{0.55}}
\put(1119.0,354.0){\rule[-0.200pt]{4.818pt}{0.400pt}}
\put(181.0,460.0){\rule[-0.200pt]{4.818pt}{0.400pt}}
\put(161,460){\makebox(0,0)[r]{0.6}}
\put(1119.0,460.0){\rule[-0.200pt]{4.818pt}{0.400pt}}
\put(181.0,566.0){\rule[-0.200pt]{4.818pt}{0.400pt}}
\put(161,566){\makebox(0,0)[r]{0.65}}
\put(1119.0,566.0){\rule[-0.200pt]{4.818pt}{0.400pt}}
\put(181.0,671.0){\rule[-0.200pt]{4.818pt}{0.400pt}}
\put(161,671){\makebox(0,0)[r]{0.7}}
\put(1119.0,671.0){\rule[-0.200pt]{4.818pt}{0.400pt}}
\put(181.0,777.0){\rule[-0.200pt]{4.818pt}{0.400pt}}
\put(161,777){\makebox(0,0)[r]{0.75}}
\put(1119.0,777.0){\rule[-0.200pt]{4.818pt}{0.400pt}}
\put(181.0,143.0){\rule[-0.200pt]{0.400pt}{4.818pt}}
\put(181,102){\makebox(0,0){0}}
\put(181.0,757.0){\rule[-0.200pt]{0.400pt}{4.818pt}}
\put(277.0,143.0){\rule[-0.200pt]{0.400pt}{4.818pt}}
\put(277,102){\makebox(0,0){0.005}}
\put(277.0,757.0){\rule[-0.200pt]{0.400pt}{4.818pt}}
\put(373.0,143.0){\rule[-0.200pt]{0.400pt}{4.818pt}}
\put(373,102){\makebox(0,0){0.01}}
\put(373.0,757.0){\rule[-0.200pt]{0.400pt}{4.818pt}}
\put(468.0,143.0){\rule[-0.200pt]{0.400pt}{4.818pt}}
\put(468,102){\makebox(0,0){0.015}}
\put(468.0,757.0){\rule[-0.200pt]{0.400pt}{4.818pt}}
\put(564.0,143.0){\rule[-0.200pt]{0.400pt}{4.818pt}}
\put(564,102){\makebox(0,0){0.02}}
\put(564.0,757.0){\rule[-0.200pt]{0.400pt}{4.818pt}}
\put(660.0,143.0){\rule[-0.200pt]{0.400pt}{4.818pt}}
\put(660,102){\makebox(0,0){0.025}}
\put(660.0,757.0){\rule[-0.200pt]{0.400pt}{4.818pt}}
\put(756.0,143.0){\rule[-0.200pt]{0.400pt}{4.818pt}}
\put(756,102){\makebox(0,0){0.03}}
\put(756.0,757.0){\rule[-0.200pt]{0.400pt}{4.818pt}}
\put(852.0,143.0){\rule[-0.200pt]{0.400pt}{4.818pt}}
\put(852,102){\makebox(0,0){0.035}}
\put(852.0,757.0){\rule[-0.200pt]{0.400pt}{4.818pt}}
\put(947.0,143.0){\rule[-0.200pt]{0.400pt}{4.818pt}}
\put(947,102){\makebox(0,0){0.04}}
\put(947.0,757.0){\rule[-0.200pt]{0.400pt}{4.818pt}}
\put(1043.0,143.0){\rule[-0.200pt]{0.400pt}{4.818pt}}
\put(1043,102){\makebox(0,0){0.045}}
\put(1043.0,757.0){\rule[-0.200pt]{0.400pt}{4.818pt}}
\put(1139.0,143.0){\rule[-0.200pt]{0.400pt}{4.818pt}}
\put(1139,102){\makebox(0,0){0.05}}
\put(1139.0,757.0){\rule[-0.200pt]{0.400pt}{4.818pt}}
\put(181.0,143.0){\rule[-0.200pt]{230.782pt}{0.400pt}}
\put(1139.0,143.0){\rule[-0.200pt]{0.400pt}{152.731pt}}
\put(181.0,777.0){\rule[-0.200pt]{230.782pt}{0.400pt}}
\put(40,460){\makebox(0,0){\rotatebox{90}{\Large\bf $m_{0^+}$ }}}
\put(660,21){\makebox(0,0){\Large\bf $m_s a$}}
\put(660,839){\makebox(0,0){\large\bf $16^3\times 72$ $m_ca = 0.431$}}
\put(181.0,143.0){\rule[-0.200pt]{0.400pt}{152.731pt}}
\put(564.0,462.0){\rule[-0.200pt]{0.400pt}{33.726pt}}
\put(554.0,462.0){\rule[-0.200pt]{4.818pt}{0.400pt}}
\put(554.0,602.0){\rule[-0.200pt]{4.818pt}{0.400pt}}
\put(660.0,494.0){\rule[-0.200pt]{0.400pt}{24.331pt}}
\put(650.0,494.0){\rule[-0.200pt]{4.818pt}{0.400pt}}
\put(650.0,595.0){\rule[-0.200pt]{4.818pt}{0.400pt}}
\put(756.0,506.0){\rule[-0.200pt]{0.400pt}{28.667pt}}
\put(746.0,506.0){\rule[-0.200pt]{4.818pt}{0.400pt}}
\put(746.0,625.0){\rule[-0.200pt]{4.818pt}{0.400pt}}
\put(852.0,513.0){\rule[-0.200pt]{0.400pt}{26.499pt}}
\put(842.0,513.0){\rule[-0.200pt]{4.818pt}{0.400pt}}
\put(842.0,623.0){\rule[-0.200pt]{4.818pt}{0.400pt}}
\put(947.0,526.0){\rule[-0.200pt]{0.400pt}{24.331pt}}
\put(937.0,526.0){\rule[-0.200pt]{4.818pt}{0.400pt}}
\put(564,532){\circle*{24}}
\put(660,545){\circle*{24}}
\put(756,566){\circle*{24}}
\put(852,568){\circle*{24}}
\put(947,576){\circle*{24}}
\put(937.0,627.0){\rule[-0.200pt]{4.818pt}{0.400pt}}
\end{picture}
\setlength{\unitlength}{0.240900pt}
\ifx\plotpoint\undefined\newsavebox{\plotpoint}\fi
\sbox{\plotpoint}{\rule[-0.200pt]{0.400pt}{0.400pt}}%
\begin{picture}(1200,900)(0,0)
\font\gnuplot=cmr10 at 10pt
\gnuplot
\sbox{\plotpoint}{\rule[-0.200pt]{0.400pt}{0.400pt}}%
\put(181.0,143.0){\rule[-0.200pt]{4.818pt}{0.400pt}}
\put(161,143){\makebox(0,0)[r]{0.45}}
\put(1119.0,143.0){\rule[-0.200pt]{4.818pt}{0.400pt}}
\put(181.0,249.0){\rule[-0.200pt]{4.818pt}{0.400pt}}
\put(161,249){\makebox(0,0)[r]{0.5}}
\put(1119.0,249.0){\rule[-0.200pt]{4.818pt}{0.400pt}}
\put(181.0,354.0){\rule[-0.200pt]{4.818pt}{0.400pt}}
\put(161,354){\makebox(0,0)[r]{0.55}}
\put(1119.0,354.0){\rule[-0.200pt]{4.818pt}{0.400pt}}
\put(181.0,460.0){\rule[-0.200pt]{4.818pt}{0.400pt}}
\put(161,460){\makebox(0,0)[r]{0.6}}
\put(1119.0,460.0){\rule[-0.200pt]{4.818pt}{0.400pt}}
\put(181.0,566.0){\rule[-0.200pt]{4.818pt}{0.400pt}}
\put(161,566){\makebox(0,0)[r]{0.65}}
\put(1119.0,566.0){\rule[-0.200pt]{4.818pt}{0.400pt}}
\put(181.0,671.0){\rule[-0.200pt]{4.818pt}{0.400pt}}
\put(161,671){\makebox(0,0)[r]{0.7}}
\put(1119.0,671.0){\rule[-0.200pt]{4.818pt}{0.400pt}}
\put(181.0,777.0){\rule[-0.200pt]{4.818pt}{0.400pt}}
\put(161,777){\makebox(0,0)[r]{0.75}}
\put(1119.0,777.0){\rule[-0.200pt]{4.818pt}{0.400pt}}
\put(181.0,143.0){\rule[-0.200pt]{0.400pt}{4.818pt}}
\put(181,102){\makebox(0,0){0}}
\put(181.0,757.0){\rule[-0.200pt]{0.400pt}{4.818pt}}
\put(277.0,143.0){\rule[-0.200pt]{0.400pt}{4.818pt}}
\put(277,102){\makebox(0,0){0.005}}
\put(277.0,757.0){\rule[-0.200pt]{0.400pt}{4.818pt}}
\put(373.0,143.0){\rule[-0.200pt]{0.400pt}{4.818pt}}
\put(373,102){\makebox(0,0){0.01}}
\put(373.0,757.0){\rule[-0.200pt]{0.400pt}{4.818pt}}
\put(468.0,143.0){\rule[-0.200pt]{0.400pt}{4.818pt}}
\put(468,102){\makebox(0,0){0.015}}
\put(468.0,757.0){\rule[-0.200pt]{0.400pt}{4.818pt}}
\put(564.0,143.0){\rule[-0.200pt]{0.400pt}{4.818pt}}
\put(564,102){\makebox(0,0){0.02}}
\put(564.0,757.0){\rule[-0.200pt]{0.400pt}{4.818pt}}
\put(660.0,143.0){\rule[-0.200pt]{0.400pt}{4.818pt}}
\put(660,102){\makebox(0,0){0.025}}
\put(660.0,757.0){\rule[-0.200pt]{0.400pt}{4.818pt}}
\put(756.0,143.0){\rule[-0.200pt]{0.400pt}{4.818pt}}
\put(756,102){\makebox(0,0){0.03}}
\put(756.0,757.0){\rule[-0.200pt]{0.400pt}{4.818pt}}
\put(852.0,143.0){\rule[-0.200pt]{0.400pt}{4.818pt}}
\put(852,102){\makebox(0,0){0.035}}
\put(852.0,757.0){\rule[-0.200pt]{0.400pt}{4.818pt}}
\put(947.0,143.0){\rule[-0.200pt]{0.400pt}{4.818pt}}
\put(947,102){\makebox(0,0){0.04}}
\put(947.0,757.0){\rule[-0.200pt]{0.400pt}{4.818pt}}
\put(1043.0,143.0){\rule[-0.200pt]{0.400pt}{4.818pt}}
\put(1043,102){\makebox(0,0){0.045}}
\put(1043.0,757.0){\rule[-0.200pt]{0.400pt}{4.818pt}}
\put(1139.0,143.0){\rule[-0.200pt]{0.400pt}{4.818pt}}
\put(1139,102){\makebox(0,0){0.05}}
\put(1139.0,757.0){\rule[-0.200pt]{0.400pt}{4.818pt}}
\put(181.0,143.0){\rule[-0.200pt]{230.782pt}{0.400pt}}
\put(1139.0,143.0){\rule[-0.200pt]{0.400pt}{152.731pt}}
\put(181.0,777.0){\rule[-0.200pt]{230.782pt}{0.400pt}}
\put(40,460){\makebox(0,0){\rotatebox{90}{\Large\bf $m_{1^-}$ }}}
\put(660,21){\makebox(0,0){\Large\bf $m_s a$}}
\put(660,839){\makebox(0,0){\large\bf $D_s^*$ on $16^3\times 72$ $m_ca = 0.431$}}
\put(181.0,143.0){\rule[-0.200pt]{0.400pt}{152.731pt}}
\put(564.0,456.0){\rule[-0.200pt]{0.400pt}{5.059pt}}
\put(554.0,456.0){\rule[-0.200pt]{4.818pt}{0.400pt}}
\put(554.0,477.0){\rule[-0.200pt]{4.818pt}{0.400pt}}
\put(660.0,472.0){\rule[-0.200pt]{0.400pt}{4.577pt}}
\put(650.0,472.0){\rule[-0.200pt]{4.818pt}{0.400pt}}
\put(650.0,491.0){\rule[-0.200pt]{4.818pt}{0.400pt}}
\put(756.0,466.0){\rule[-0.200pt]{0.400pt}{4.095pt}}
\put(746.0,466.0){\rule[-0.200pt]{4.818pt}{0.400pt}}
\put(746.0,483.0){\rule[-0.200pt]{4.818pt}{0.400pt}}
\put(852.0,472.0){\rule[-0.200pt]{0.400pt}{4.095pt}}
\put(842.0,472.0){\rule[-0.200pt]{4.818pt}{0.400pt}}
\put(842.0,489.0){\rule[-0.200pt]{4.818pt}{0.400pt}}
\put(947.0,478.0){\rule[-0.200pt]{0.400pt}{3.854pt}}
\put(937.0,478.0){\rule[-0.200pt]{4.818pt}{0.400pt}}
\put(564,467){\circle*{12}}
\put(660,481){\circle*{12}}
\put(756,475){\circle*{12}}
\put(852,480){\circle*{12}}
\put(947,486){\circle*{12}}
\put(937.0,494.0){\rule[-0.200pt]{4.818pt}{0.400pt}}
\end{picture}
\end{center}
\end{figure}

We then fix the heavy quark at the charm mass and fit the light quark
masses which correspond to the strange mass at $m_s a=0.0205$. This way, we obtained
the charmed strange meson masses for each of the meson interpolation field. 
The numerical results are shown in Table 1. The experimental data are  from the
PDG particle listings.

Our lattice results are plotted in Fig. 5 together with the experimental data.

\newpage
\clearpage
\begin{table}[t]
\begin{center}
\caption{
 charmed strange meson masses }
\begin{tabular}{|c|c|c|c|}
\hline
Particle &Mass( $\times a$ )  & Lattice (MeV)&Exp. (MeV) \\
\hline
$D_s(0^-)$       & 0.5608(31) & 1976(11) &1968.49(34)\\
\hline
$D_s^*(1^-)$     & 0.6049(36) & 2131(13) &2112.3(5)\\
\hline
$D_{s0}^*(0^+)$  & 0.638(22)  & 2248(78) &2317.8(6)\\
\hline
$D_{s1}(1^+_a)$    & 0.684(18)  & 2410(63) &2459.6(6)\\
\hline
$D_{s1}(1^+_b)$  & 0.703(26)  & 2476(92) &2535.35(84)\\
\hline
\end{tabular}
\end{center}
\end{table}
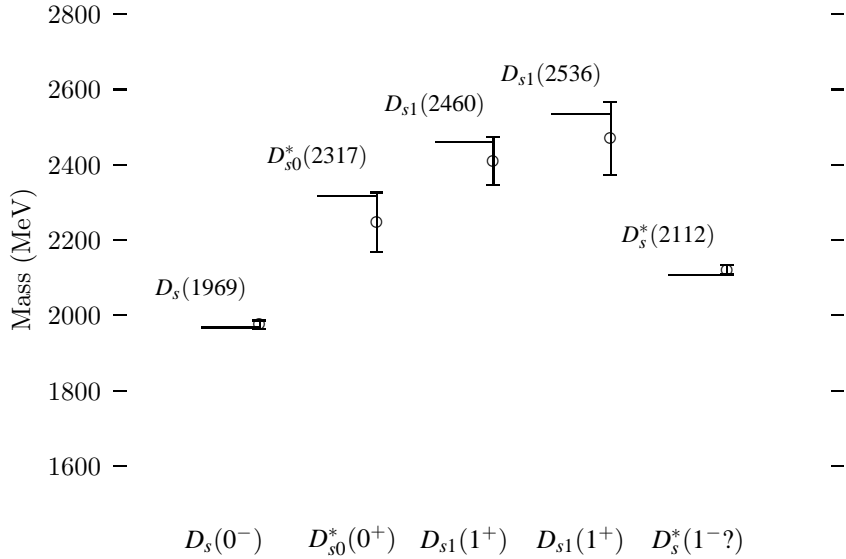
\begin{figure}[th]
\begin{center}
\caption{\bf Calculated charmed strange meson spectrum in comparison with the experimental
results}
\vspace*{0.4cm}
\setlength{\unitlength}{0.240900pt}
\ifx\plotpoint\undefined\newsavebox{\plotpoint}\fi
\sbox{\plotpoint}{\rule[-0.200pt]{0.400pt}{0.400pt}}%
\begin{picture}(1349,809)(0,0)
\font\gnuplot=cmr10 at 10pt
\gnuplot
\sbox{\plotpoint}{\rule[-0.200pt]{0.400pt}{0.400pt}}%
\put(181.0,99.0){\rule[-0.200pt]{4.818pt}{0.400pt}}
\put(161,99){\makebox(0,0)[r]{1600}}
\put(1308.0,99.0){\rule[-0.200pt]{4.818pt}{0.400pt}}
\put(181.0,217.0){\rule[-0.200pt]{4.818pt}{0.400pt}}
\put(161,217){\makebox(0,0)[r]{1800}}
\put(1308.0,217.0){\rule[-0.200pt]{4.818pt}{0.400pt}}
\put(181.0,336.0){\rule[-0.200pt]{4.818pt}{0.400pt}}
\put(161,336){\makebox(0,0)[r]{2000}}
\put(1308.0,336.0){\rule[-0.200pt]{4.818pt}{0.400pt}}
\put(181.0,454.0){\rule[-0.200pt]{4.818pt}{0.400pt}}
\put(161,454){\makebox(0,0)[r]{2200}}
\put(1308.0,454.0){\rule[-0.200pt]{4.818pt}{0.400pt}}
\put(181.0,572.0){\rule[-0.200pt]{4.818pt}{0.400pt}}
\put(161,572){\makebox(0,0)[r]{2400}}
\put(1308.0,572.0){\rule[-0.200pt]{4.818pt}{0.400pt}}
\put(181.0,691.0){\rule[-0.200pt]{4.818pt}{0.400pt}}
\put(161,691){\makebox(0,0)[r]{2600}}
\put(1308.0,691.0){\rule[-0.200pt]{4.818pt}{0.400pt}}
\put(181.0,809.0){\rule[-0.200pt]{4.818pt}{0.400pt}}
\put(161,809){\makebox(0,0)[r]{2800}}
\put(40,424){\makebox(0,0){\rotatebox{90}{Mass (MeV)}}}
\put(355,-18){\makebox(0,0){\small $D_s(0^{-})$}}
\put(557,-18){\makebox(0,0){\small $D_{s0}^*(0^+)$}}
\put(732,-18){\makebox(0,0){\small $D_{s1}(1^+)$}}
\put(915,-18){\makebox(0,0){\small $D_{s1}(1^+)$}}
\put(1099,-18){\makebox(0,0){\small $D_s^*(1^-?)$}}
\put(319,377){\makebox(0,0){\footnotesize $D_{s}(1969)$}}
\put(686,667){\makebox(0,0){\footnotesize $D_{s1}(2460)$}}
\put(869,714){\makebox(0,0){\footnotesize $D_{s1}(2536)$}}
\put(502,584){\makebox(0,0){\footnotesize $D_{s0}^*(2317)$}}
\put(1053,461){\makebox(0,0){\footnotesize $D_{s}^*(2112)$}}
\put(1308.0,809.0){\rule[-0.200pt]{4.818pt}{0.400pt}}
\put(319,317){\usebox{\plotpoint}}
\put(319.0,317.0){\rule[-0.200pt]{21.922pt}{0.400pt}}
\put(502,524){\usebox{\plotpoint}}
\put(502.0,524.0){\rule[-0.200pt]{22.163pt}{0.400pt}}
\put(686,608){\usebox{\plotpoint}}
\put(686.0,608.0){\rule[-0.200pt]{21.922pt}{0.400pt}}
\put(869,652){\usebox{\plotpoint}}
\put(869.0,652.0){\rule[-0.200pt]{22.163pt}{0.400pt}}
\put(1053,399){\usebox{\plotpoint}}
\put(1053.0,399.0){\rule[-0.200pt]{21.922pt}{0.400pt}}
\put(410.0,315.0){\rule[-0.200pt]{0.400pt}{3.132pt}}
\put(400.0,315.0){\rule[-0.200pt]{4.818pt}{0.400pt}}
\put(400.0,328.0){\rule[-0.200pt]{4.818pt}{0.400pt}}
\put(594.0,436.0){\rule[-0.200pt]{0.400pt}{22.404pt}}
\put(584.0,436.0){\rule[-0.200pt]{4.818pt}{0.400pt}}
\put(584.0,529.0){\rule[-0.200pt]{4.818pt}{0.400pt}}
\put(777.0,541.0){\rule[-0.200pt]{0.400pt}{18.067pt}}
\put(767.0,541.0){\rule[-0.200pt]{4.818pt}{0.400pt}}
\put(767.0,616.0){\rule[-0.200pt]{4.818pt}{0.400pt}}
\put(961.0,556.0){\rule[-0.200pt]{0.400pt}{27.703pt}}
\put(951.0,556.0){\rule[-0.200pt]{4.818pt}{0.400pt}}
\put(951.0,671.0){\rule[-0.200pt]{4.818pt}{0.400pt}}
\put(1144.0,400.0){\rule[-0.200pt]{0.400pt}{3.613pt}}
\put(1134.0,400.0){\rule[-0.200pt]{4.818pt}{0.400pt}}
\put(410,322){\circle{18}}
\put(594,482){\circle{18}}
\put(777,578){\circle{18}}
\put(961,614){\circle{18}}
\put(1144,407){\circle{18}}
\put(1134.0,415.0){\rule[-0.200pt]{4.818pt}{0.400pt}}
\end{picture}
\end{center}
\end{figure}

   We see that the pseudoscalar ($D_s$) and the vector ($D_s^*$) agree with
the experimental results very well, assuming the $D_s^*$ at 2112 MeV is the
$1^-$ meson. All the p-waves states ($0^+$ and the two $1^+$) agree with
the experiments within one sigma. In particular, we find the 
$0^+$ meson with the $\overline{\psi}\psi$ interpolation field at 2248(78) MeV to be
much higher than the quark model prediction. This suggests that $D_{s0}^*(2317)$
may well be the conventional $c\bar{s}$ meson.

To have a better estimate of the $D_{s0}^*$ mass, we used the ratio method to calculate 
the mass difference. Since the
correlators are calculated from the same gauge configurations, the ratio
method is expected to reduce fluctuations. We calculated the ratio between
the scalar and pseudoscalar correlators.
\begin{eqnarray}
\frac{G(0^+)}{G(0^-)} &\sim & exp(-\Delta m\cdot t), \,\,\,\, ~~~\Delta m = m(0^+)-m(0^-).
\end{eqnarray}
The preliminary result is $\Delta m = m(D_{s0}^*)-m(D_s)=275(53)$ MeV which
is about a sigma from the experimental result of 349.3(7) MeV.

 To summarize, we have carried out a calculation of the charmed strange meson
masses with the overlap fermion on quenched $16^3 \times 72$ lattices. The calculated masses are 
consistent with the experimental results within error bars. The $J^p=1^-$ meson
mass matches with $D_s^*(?)(2112)$ very well. This implies that $D_s^*(?)(2112)$ 
is indeed the expected vector meson. The scalar meson mass at 2248(78) MeV is
higher than the prediction of the quark model and is in agreement with the experimental
mass of $D_{s0}^* (2317)$.

This work is partially supported by DOE Grands DE-FG05-84ER40154. We thank 
H.Y. Cheng for bringing the problem of $D_{s0}^* (2317)$ to our attention.

\end{document}